# Real-space Observation of Incommensurate Spin Density Wave and Coexisting Charge Density Wave on Cr (001) surface

Yining Hu[1]†, Tianzhen Zhang[1]†, Dongming Zhao[1]†, Chen Chen[1], Shuyue Ding[1], Wentao Yang[1], Xu Wang[1], Chihao Li[1], Haitao Wang[1], Donglai Feng[2,3,4], Tong Zhang[1,3,4]*

[1] State Key Laboratory of Surface Physics, Department of Physics, and Advanced Materials Laboratory, Fudan University, Shanghai 200438, China
[2] Hefei National Laboratory for Physical Science at Microscale and Department of Physics, University of Science and Technology of China, Hefei, Anhui 230026, China
[3] Collaborative Innovation Center of Advanced Microstructures, Nanjing 210093, China
[4] Shanghai Research Center for Quantum Sciences, Shanghai 201315, China

† These authors contributed equally.
*Email: tzhang18@fudan.edu.cn

In itinerant magnetic systems, a spin density wave (SDW) state can be induced by Fermi surface nesting and electron-electron interaction. It may intertwine with other orders such as charge density wave (CDW), while their relation is still yet to be understood. Here via spin-polarized scanning tunneling microscopy, we directly observed long-range spin modulation on Cr(001) surface, which corresponds to the well-known incommensurate SDW of bulk Cr. It displays 6.0 nm in-plane period and anti-phase behavior between adjacent (001) planes. Meanwhile, we simultaneously observed the coexisting CDW with half the period of SDW. Such SDW/CDW have highly correlated domain structures and are in-phase. Surprisingly, the CDW displays a contrast inversion around a density-of-states dip at -22 meV, indicating an anomalous CDW gap opened below $E_F$. These observations support that the CDW is a secondary order driven by SDW. Our work is not only a real-space characterization of incommensurate SDW, but also provides insights on how SDW and CDW coexist.

A spin density wave (SDW) state manifests itself as real-space spin modulations. It is usually formed in itinerant magnetic systems with Fermi surface nesting and electron-electron interactions[1]. The spatial period of SDW could be commensurate (C-SDW) or incommensurate (IC-SDW) to lattice constant. In the latter case, the spin modulation decouples from lattice, which is distinguished from local moment induced anti-ferromagnetic (AFM) order. Interestingly, SDW often coexists and sometimes intertwines with other orders in correlated systems, such as charge density wave (CDW) and superconductivity[1-4]. The interplay of these coexisting/intertwining orders has now become an important theme in condensed matter physics. To date, the commonly observed SDW states are commensurate SDW, such as the collinear/bicollinear SDW (AFM) state in iron-based superconductors[4,5]; while incommensurate SDW is rarely seen, and particularly, its real space imaging is quite lacking.

Chromium (Cr) is one of the classic examples which shows itinerant magnetism with an IC-SDW ground state[6-8] below its Néel temperature ($T_N$ = 311 K). Such IC-SDW is stabilized by "imperfect" Fermi surface nesting condition[9], as illustrated in Fig. 1a. Specifically, the Fermi surfaces of Cr (b.c.c. lattice) are composed of hole pockets at the corner and electron pocket at the center of the Brillouin zone[6]. The hole pocket is slightly larger than the electron pocket which yields two nesting vectors: $\mathbf{Q}_\pm = 2\pi/a\,(1 \pm \delta)$ ($a$ = 2.9 Å being the lattice constant). Therefore, a long period IC-SDW with a wave vector $\mathbf{Q}_{SDW} = 2\pi\delta/a$ is generated which overlaps with the AFM coupling between Cr atoms (Fig. 1b). The wavelength of IC-SDW is reported to be 6.0 nm at $T$ < 10 K (ref. 10), and $\mathbf{Q}_{SDW}$ is along one of the <001> directions. The spin orientation of Cr atom is found to be perpendicular to $\mathbf{Q}_{SDW}$ at $T > T_{SF}$ (123 K) but switched to be parallel at $T < T_{SF}$ (spin-flip transition[6]).

In addition to IC-SDW, a charge density wave (CDW) with half period of the IC-SDW was also found in Cr [10~13]. Unlike the IC-SDW, the exact origin of such CDW is yet to be understood. It was often considered as the second-order harmonics of IC-SDW[12], corresponding to a nesting vector $\mathbf{Q}_{CDW} = 2\mathbf{Q}_{SDW}$ that connects the two folded bands at Γ (Fig. 1a); alternatively, it was suggested as a lattice strain wave induced by magneto-elastic coupling to the IC-SDW[6]. Therefore, being a pure element arranged in a simple structure, Cr is also a classical system to study the interplay of SDW/CDW orders.

However, after decades of research, the characterization of IC-SDW (and CDW) in Cr is still rather limited to spatially averaged method, such as neutron scattering[6,7], x-ray diffraction[10,13] and photoemission spectroscopy[14,15]. In principle, SDW could also be detected by local probes at atomic scale, such as spin-polarized scanning tunneling microscopy (SP-STM)[16]. Although a few SP-STM studies have been performed on various Cr surfaces[17-26], the real-space evidence of IC-SDW was rarely reported (some studies found CDW modulation on Cr (110) surface[23, 27], and argued the satellite FFT spots as an indication of IC-SDW[23]). Most SP-STM studies on Cr (001) surface only observed in-plane ferromagnetism with AFM coupling between adjacent (001) planes[17-22]. To understand such a ferromagnetic spin arrangement on surface, it was suggested the magnetic moment is enhanced at the surface[28] and the IC-SDW antinodes are always pinned on the surface[8,19], making it invisible to STM. However, we noticed that most previous STM studies on Cr (001) did not resolve clear atomic lattices, although mono-atomic terrace can be identified. This is likely due to local disorders induced by segregated impurities on the surface, which is a common problem in cleaning Cr single crystal. As the surface conditions could alter the surface magnetism[8], it would be intriguing to search the IC-SDW in real space again on a well-ordered Cr surface.

In this work, by using low-temperature spin-polarized STM with vector magnetic field, we studied a thoroughly cleaned Cr (001) with a well-ordered surface. We observed clear spin modulation with a period of 6.0 nm, propagating along in-plane [100] or [010] directions, which well matches the projected bulk IC-SDW on (001) surface. Its SDW nature is confirmed by the contrast inversion upon switching tip's magnetization, and the anti-phase relation between adjacent terraces. Meanwhile, we also observed the coexisting CDW with a period 3.0 nm, and surprisingly found that it displays a π phase shift around gap structure about 22 meV below $E_F$, which suggests its formation is beyond the intuitive Fermi surface nesting picture. Furthermore, as a local probe measurement, we directly observed the domain structure of SDW/CDW and revealed their in-phase relation. Our work not only gives a real-space investigation of IC-SDW, but also provide new insights on the general mechanism of coexisting SDW/CDW orders.

**Results**

**STM characterization of Cr (001) surface and it's tunneling spectrum.**

The experiment was conducted in a cryogenic STM (UNISOKU) at $T$ = 5.0 K. Details about the cleaning process of Cr (001) and STM measurement are described in the Method section. Fig. 1c shows a large scale STM image of the obtained Cr (001) surface. It displays atomically flat terraces with mono-atomic height (≈ 0.14 nm), as indicated by the line profile in Fig. 1e. It is notable that the terrace edges prefer running along high symmetric directions such as [100], [010] and [110]. This is an indication of free surface atom diffusion during annealing[29]. Despite some randomly distributed defects, atomic lattice can be easily resolved in defect free area, as shown in Fig. 1d. It displays a centered 2×2 (or √2×√2 R45°) lattice with respect to the pristine Cr 1×1 lattice. Some dislocation lines are observed where the atomic lattice on their two sides displays certain shift (black dashed lines in Fig. 1d). These dislocations do not show influence to SDW/CDW discussed below, more details are presented in Supplementary Fig. S1. We noticed some previous STM works on Cr (001) also observed c(2×2) structures[22,26,30], but the electronic states of the present surface is quite different from those studies (shown below). Although the origin of this reconstruction is unclear at this stage, it is the first time to observe regular oriented terrace edges with well-ordered lattice on a sputtered/annealed Cr (001) surface.

The typical large energy scale d$I$/d$V$ spectrum of the surface, measured by a normal W tip above defect-free area, is shown in Fig. 1f. There is a pronounced DOS peak located at -75 (±5) mV. We note although a DOS peak was widely observed on Cr surfaces[18-26, 31-32], the peak position varies significantly for different studies. The origin of such a peak was usually attributed to spin-polarized surface state[31,33-35] or the orbital Kondo effect[32]. Our measurements shown below tend to support the former scenario. By zooming into a narrower energy range near $E_F$ (Fig. 1g), we found there is an additional DOS dip at $E$ = -22(±1) meV, which has never been reported before. Such a DOS dip is repeatedly observed at different surface locations (see Supplementary Fig. S2 for more spectra). We will show later that it is likely an energy gap associated with the CDW order, but opens below the Fermi level.

**Real-space imaging of incommensurate SDW.**

We then studied the surface with spin-polarized tips. Fig. 2a is a d$I$/d$V$ map taken at $V_b$ = -150 meV with a tip coated with 40 nm thick Cr, which favors an in-plane spin polarization[16]. The mapping area is the same as that shown Fig. 1c. It is remarkable that stripe-like modulations can be observed, and there are two domains of such modulation which are perpendicular to each other. A zoomed-in d$I$/d$V$ map around a domain wall is shown in Fig. 2b. The period of the stripe is 6.0 nm and the wave vector is either along [010] or [100] direction (as also seen in the FFT image in Fig. 2a inset). Such a period and propagating direction exactly match the projected bulk IC-SDW of Cr on a (001) surface. It is also seen that the domain walls in Fig. 2a (dashed curve) have no correlation with surface morphology (Fig. 1c), which indicates the stripes are not merely surface effects but of bulk origin. Fig. 2c shows the typical d$I$/d$V$ spectra taken on the stripe and between the stripes with the Cr-coated tip. There is observable difference on the intensity around the DOS peak, which is attributed to spin contrast as discussed below. More d$I$/d$V$ maps taken at different energies and their FFT images can be found in Supplementary Figs. S3 and S4.

To further verify these stripes are spin modulations, we performed measurement with a 16 nm-Fe-coated tip whose magnetization can be controlled by an external magnetic field[16]. Figs. 3a and 3b are two dI/dV maps taken in the same region, but under opposite in-plane field of $B_X$ = ± 1T (**X** direction is perpendicular to stripes, as marked in figure). The 6.0 nm period stripes can be seen in both Figs. 3a and 3b, while they display a clear phase inversion, as further illustrated in their line profiles in Fig. 3e. Since the tip magnetization will follow such in-plane field, this gives a direct evidence that the stripes are SDW modulations, with opposite spin orientations on their peaks and troughs. We can also tune the tip magnetization along **Y** and **Z** directions (by applying $B_Y$ = 1T and $B_Z$ = 1.5T, respectively), the resulting d$I$/d$V$ maps are shown in Figs. 3c and 3d, respectively. In these two cases, the contrast of the stripes was significantly reduced and almost invisible. This confirmed that the spins are only polarized along **X** direction (the same direction of $\mathbf{Q}_{SDW}$), which agrees with bulk measurements that the IC-SDW is longitudinal wave at $T < T_{SF}$ (ref. 6). We can extract the spin-polarization ratio (spin-contrast) of Figs. 3a and 3b by calculating their relative intensity difference, which is about 4% at the SDW peaks (Fig. 3f). We note here that for Cr-coated tip used in Fig. 2, its (in-plane) polarization direction is arbitrary[16], that is why the two SDW domains in Figs. 2a, 2b can be simultaneously imaged but their contrasts are different.

We note that the above measurement under vector magnetic field also distinguished the SDW state from "spin-spiral" order which has been detected by SP-STM in other magnetic systems[36,37]. In a spin-spiral, the spins have a nearly constant magnitude but their orientations keep rotating with certain chirality, thus one would observe spin modulation in at least two of the X, Y, Z components (depends on the types of spin-spiral, e.g., helical or cycloidal[36]). However here we only observed spin modulation along X direction.

Another signature of IC-SDW can be obtained near the atomic step edges. Fig. 3g shows a topographic image of three adjacent mono-atomic height terraces, and Fig. 3h is the corresponding dI/dV map measured by a Cr-coated tip. As shown by dashed lines, there is a phase inversion of the modulations between adjacent terraces. This indicates the local spin of two adjacent (001) plane are still AFM coupled. Based on above observations, we now achieve a complete spin configuration of the present Cr (001) surface. As illustrated in Fig. 3i, the Cr have an out-of-plane AFM configuration with the spins lying in-plane, while a long wavelength, longitudinal IC-SDW ($\lambda$ = 6.0 nm) is present in each (001) planes. Such a magnetic structure agrees with the neutron-scattering measurement for bulk Cr and thick Cr films[6-8], but has not been visualized by a local probe before. Comparing with the commensurate SDW or AFM state[4,16], the IC-SDW observed here are pure spin modulations that decoupled from the lattice. It can be considered as a "modulated" ferromagnetism for the top Cr plane.

**Observation of the coexisting CDW.**

Moreover, in addition to the SDW modulation, we also observed another type of modulation with half the period of SDW (3.0 nm), as shown in the dI/dV map in Fig. 2d (taken at $V_b$ = -10 mV). It displays the same domain structures with the SDW shown in Fig. 2b, however here the two domains have the same contrast. We further verified that such 3.0 nm modulation is also visible under a nonmagnetic PtIr tip, but the SDW is invisible (see Supplementary Fig. S5). Therefore, it is natural to assign such spin-unpolarized modulation to CDW with a $\mathbf{Q}_{CDW}$ = 2$\mathbf{Q}_{SDW}$, as reported in X-ray studies of Cr[10,13]. We noticed a previous STM study on Cr (110) surface reported similar charge modulation that originated from bulk CDW[27]. Here we are able to image the SDW and CDW simultaneously, enabling the study of their microscopic correlations.

Figs. 4a-h show a series of dI/dV maps taken in the same region, with the same Cr tip but at various $V_b$. Their (averaged) line profiles are summarized in Fig. 4j, which display the evolvement of SDW/CDW modulations as the energy varies. Fig. 4i shows the typical dI/dV spectrum of this mapping region, and the energy positions corresponding to line profiles in Fig. 4j are indicated by dashed lines. The SDW modulation is mainly observable in the energy range of -200 meV ~ -50 meV, which covers the large DOS peak in dI/dV. This suggests the peak is from certain spin-polarized state(s). As the mapping energy lowered to -50 meV ~ 0, the 3.0 nm CDW modulation became pronounced. Interestingly, it displays an abrupt phase inversion between -30 meV and -10 meV (see also the dI/dV maps in Figs. 4f and 4g). Such π phase shift can also be seen in the phase of the Fourier transformations, as shown in Fig. 4k. We note this energy range right covers the DOS dip in the dI/dV spectrum (Fig. 4i), which suggests such a DOS dip is from a CDW gap, as the phase inversion of particle-hole states around the gap is a hallmark of CDW[38,39]. Our measurement at elevated temperatures (shown in Supplementary Figs. S6 and S7) also suggested the DOS dip and CDW are correlated, as they are still both visible at $T$ = 78 K but disappeared together at $T$ = 301 K, which is close to $T_N$ = 311 K.

However, it is unusual that the gap is not opened at $E_F$ here, but about 22 meV below, which is rather unexpected for conventional CDW [1]. This gap is also unlikely associated with the SDW as its size (≈ 10 meV) is too small with comparing to the Néel temperature of Cr (311 K). We note in previous ARPES study on Cr (110), a SDW gap is found to be about 200 meV and located above $E_F$ (ref.14), which can be understood through AFM coupling induced band folding (see Supplementary Fig. S8, the hole pocket of Cr is slightly larger than electron pocket, their crossing point upon folding is above $E_F$). However, here we did not observe an obvious SDW gap in tunneling spectrum, although the tunneling conductance at positive energy is indeed low in Fig. 1f (whether it is related to SDW gap needs further investigation). Assuming the Cr sample here has similar band structure to that reported in ref. 14, a CDW gap below $E_F$ cannot be induced by the band folding scenario either. It therefore suggests the formation of CDW is beyond the intuitive Fermi surface nesting picture. We note a recent STM study on TiSe$_2$ also indicated a CDW gap opened away from $E_F$ (ref. 40), which was attributed to strong electron correlations.

Further information can be extracted from the real space imaging of SDW and CDW is their phase relation. As discussed before, the SDW modulations in d$I$/d$V$ are directly induced by spin contrast (Fig. 3). Their maximum and minimum positions are where the absolute spin density reach maximum. Meanwhile, the CDW modulation in d$I$/d$V$ is the local DOS variation induced by periodic charge distribution[38]. Fig. 4j shows that the locations with maximum spin density (tracked by solid and dashed lines) always have minimum LDOS at -30 mV and maximum LDOS at -10 meV. As usually the charge density is proportional to the LDOS of occupied state near $E_F$, our data suggests the SDW and CDW in Cr are in-phase, i.e., the positions with maximum spin density also have maximum charge density (sketched in Fig. 1b). We note that previously such relation can only be obtained through combined X-ray and neutron diffraction measurement after extensive data analysis[41,42], while here we provide a rather direct evidence on the same system. The in-phase relation appears to consist with the theory which treat CDW as a second harmonics of SDW[12,43,44].

At last, beside the static SDW/CDW modulations, we also observed dispersive quasi-particle interference (QPI) on the surface. As shown in Fig. 5a for example, clear short wavelength interference patterns are visible around the defects. Fig. 5b is the FFT image of Fig. 5a which displays a square shaped scattering ring (more QPI data is shown in Supplementary Fig. S9). By summarizing the FFT line profile taken at various energies (Fig. 5c), an electron-like dispersion with $q_F \approx 1.1$ Å$^{-1}$ is visualized. We note previous DFT calculations had predicted multiple spin-polarized surface states on Cr(001)[33], the observed QPI could be originated from one of the surface states, as bulk bands usually do not generate strong QPI; and the DOS peak at $E$ = -75 meV in d$I$/d$V$ (Fig. 1f) could be from the onset of this band. The clear observation of QPI here (which is absent in previous STM studies) is also an indication of improved surface condition. We expect it will help to elucidate the surface electronic states of Cr with the help of further theoretical calculations.

**Discussion**

We have now presented a comprehensive SP-STM study on a well-ordered Cr (001) surface. We directly observed the incommensurate SDW with similar spin configurations to the bulk, which manifests as long-period, linearly polarized pure spin modulations on the surface; and the coexisting CDW order is also simultaneously observed. These features are absent in previous STM studies which indicates the surface condition is important for probing intrinsic magnetism of Cr. Another main finding of this work is the CDW gap opened below $E_F$. It appears conflict with conventional Fermi surface nesting picture in which the density wave gap will open at $E_F$ to lower the energy. Nonetheless, if considering the CDW in Cr is driven by SDW[12,43,44], the system would gain energy mainly through SDW and it is possible the formation of CDW involves states away from $E_F$. This implies additional factors, such as electron correlations, may play a role in CDW of Cr. Moreover, as a real space measurement, we can directly observe the domain and phase relations of SDW and CDW. The same domain structure and an in-phase relation of local spin and charge indicate these two orders are highly correlated, consistent with the scenario that CDW is the high-order harmonics of SDW.

The mechanism of coexisting spin/charge orders has long been an important issue in condensed matter physics, particularly for correlated materials such as cuprates[3,45] and iron-based superconductors[4,46], our new spectroscopic and microscopic information provide insights on the comprehensive understanding SDW/CDW in Cr and other correlated materials. Our work is one of the few cases in which simultaneous imaging of spin/charge order with high resolution is achieved (ref. 5 is another example). We expected similar SP-STM measurement shall also be applied to other systems and would inspire more studies on the coexisting quantum orders.

**Methods**

**Cr (001) sample preparation and STM measurements.** Cr (001) single crystal (Mateck, purity: 99.999%) was intensively cleaned by repeated cycles of Ar sputtering at 750°C (for 15 min) and annealing at 800°C (for 20 min), until a well-ordered surface is obtained. Spin-resolved tunneling spectroscopy and conductance mapping were performed by Cr and Fe coated STM tips, which are prepared by depositing 40 nm Cr or 16 nm Fe layers on W tip. The W tip was electrochemically etched and flashed up to ≈ 2000 K for cleaning before coating. The tunneling conductance (d$I$/d$V$) was collected by standard lock-in method and the bias voltage ($V_b$) is applied to the sample.

**Data availability**

The main data supporting the findings of this study are available within the article and its Supplementary Information files. All the raw data generated in this study are available from the corresponding author upon reasonable request.

**Code availability**

All the data analysis codes related to this study are available from the corresponding author upon reasonable request.

## Acknowledgments

We thank Prof. Chunlei Gao and Mr. Zhongjie Wang for the advice of preparing spin-polarized tip and helpful discussions. This work is supported by the National Key R&D Program of the MOST of China with Grant Nos. 2017YFA0303004 (T. Z.), National Natural Science Foundation of China with Grant Nos. 92065202 (T.Z.), 11888101 (D.L.F.), 11790312 (D.L.F.), 11961160717 (T.Z.), Science Challenge Project with grant No. TZ2016004 (D.L.F.), Shanghai Municipal Science and Technology Major Project with Grant No. 2019SHZDZX01 (T.Z., D.L.F.), Science and Technology Commission of Shanghai Municipality, China (Grant No. 19JC1412702 (T.Z.)).

## Author contributions

The STM measurements and data analysis were performed by Y. N. Hu, T. Z. Zhang, D. M. Zhao, C. Chen, S. Y. Ding, W. T. Yang, X. Wang, C.H. Li, H. T. Wang and T. Zhang. D. L. Feng and T. Zhang coordinated the project and wrote the manuscript. All authors have discussed the results and the interpretation.

## Competing interests

The authors declare no competing interests.

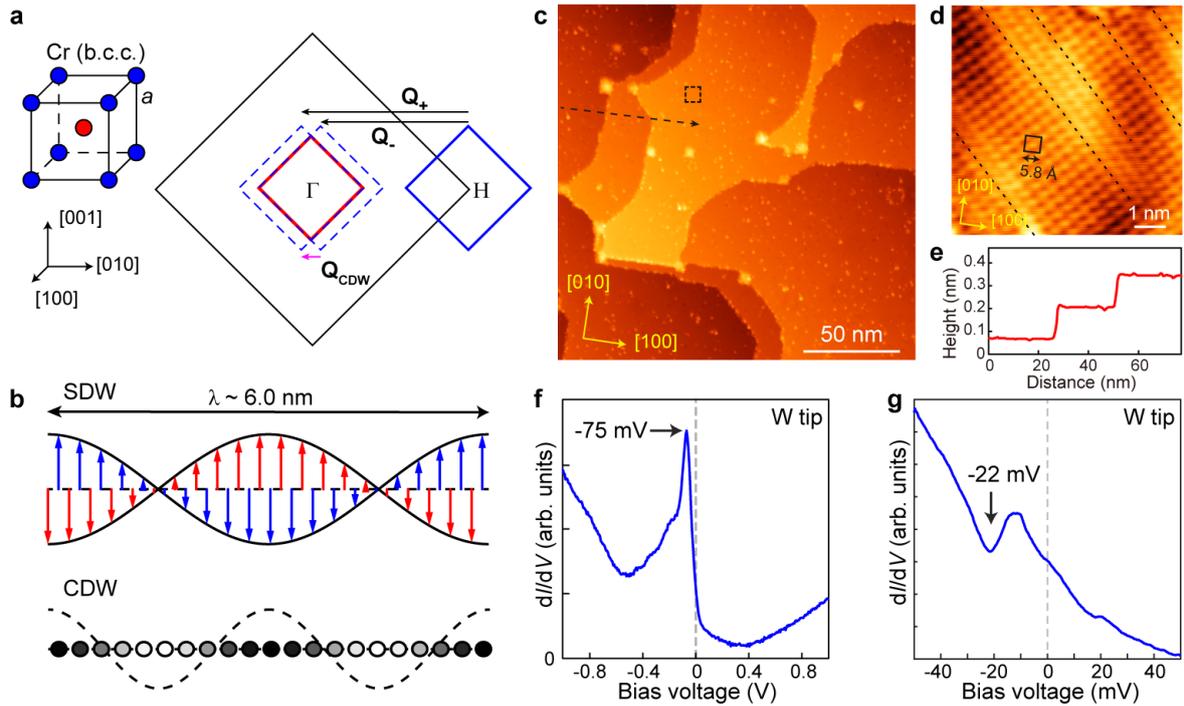

**Fig. 1 Sketch of the crystal structure and SDW/CDW states of bulk Cr, and STM characterization of Cr(001) surface. a** Left: the body-centered cubic (b.c.c.) structure, and right: the (001) plane Brillouin zone of Cr. Electron and hole Fermi surfaces (cross sections) are represented by red and blue squares, respectively. The nesting vectors $\mathbf{Q}_\pm = 2\pi/a\,(1 \pm \delta)$ are indicated by black arrows, and $\mathbf{Q}_{CDW} = \mathbf{Q}_+ - \mathbf{Q}_-$ is the CDW wave vector. **b** Schematic of incommensurate SDW and CDW of bulk Cr in real space. Blue and red arrows represent the spin of corner and body-center Cr atoms in the b.c.c. lattice at $T > T_{SF}$, respectively. Solid (hollow) circles represent the locations with the highest (lowest) charge density. **c** Large scale STM image (190×190 nm$^2$) of the cleaned Cr (001) surface. **d** Atomically resolved STM image (taken in the dashed square in panel (**c**)) showing a centered (2×2) structure. The dashed lines indicate the surface line-dislocations (see Supplementary Fig. S1 for more details). **e** Line profile taken along the dashed line in panel (**c**). **f** Typical d$I$/d$V$ point spectrum taken on a defect-free region with a normal W tip (setpoint: $V_b$ = 1 V, $I$ = 60 pA, Δ$V$ = 20 mV). **g** Typical d$I$/d$V$ spectrum taken around $E_F$ by W tip (setpoint: $V_b$ = 50 mV, $I$ = 100 pA, Δ$V$ = 1 mV), a DOS dip at $V_b$ ≈ -22 (±1) mV is observed.

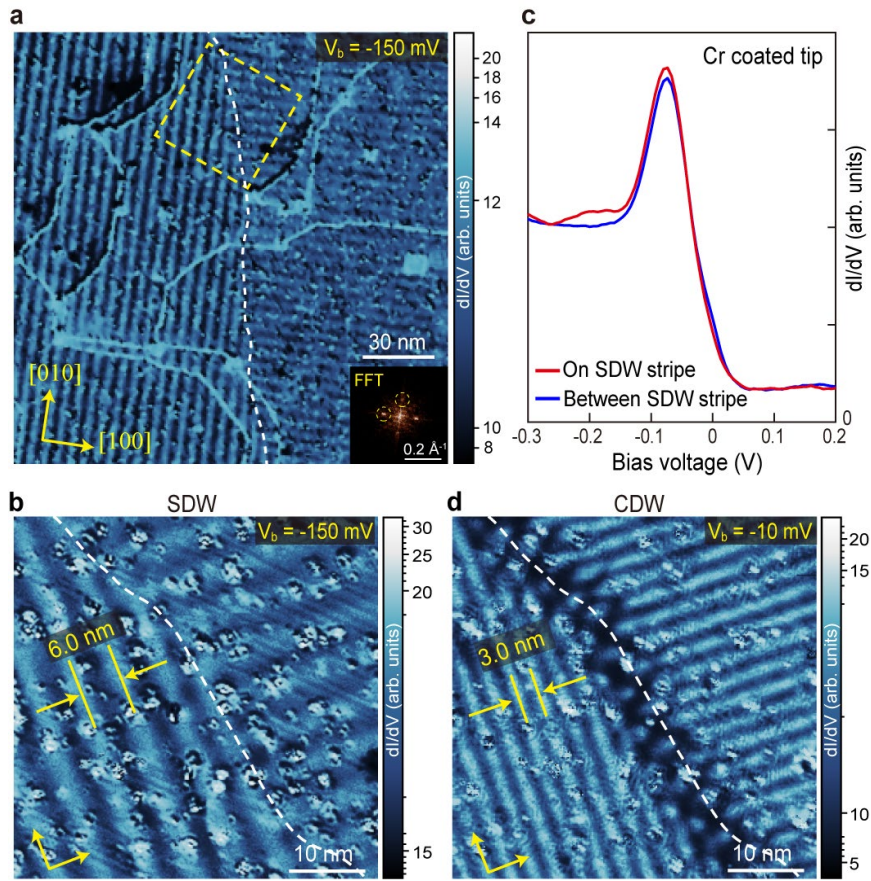

**Fig. 2 Spin and charge modulations observed on Cr (001) surface. a** The d*I*/d*V* map of the same area that shown in Fig. 1c, taken by a Cr coated spin-polarized tip at $V_b$ = -150 mV (*I* = 80 pA, Δ*V* = 20 mV). Inset: FFT image, dashed circles indicate the spots from spin modulations. **b** A zoom-in d*I*/d*V* map of the marked square in panel (**a**), which shows spin modulation with a period of 6.0 nm. White dashed curve tracks the domain wall ($V_b$ = -150 mV, *I* = 150 pA, Δ*V* = 15 mV). **c** Typical d*I*/d*V* (point spectra) taken on and in between the spin modulation stripes (setpoint: $V_b$ = 1 V, *I* = 60 pA, Δ*V* = 20 mV). **d** d*I*/d*V* map of the same area as panel (**b**), but taken at $V_b$ = -10 mV. The charge modulation with a period of 3.0 nm is observed (*I* = 150 pA, Δ*V* = 5 mV).

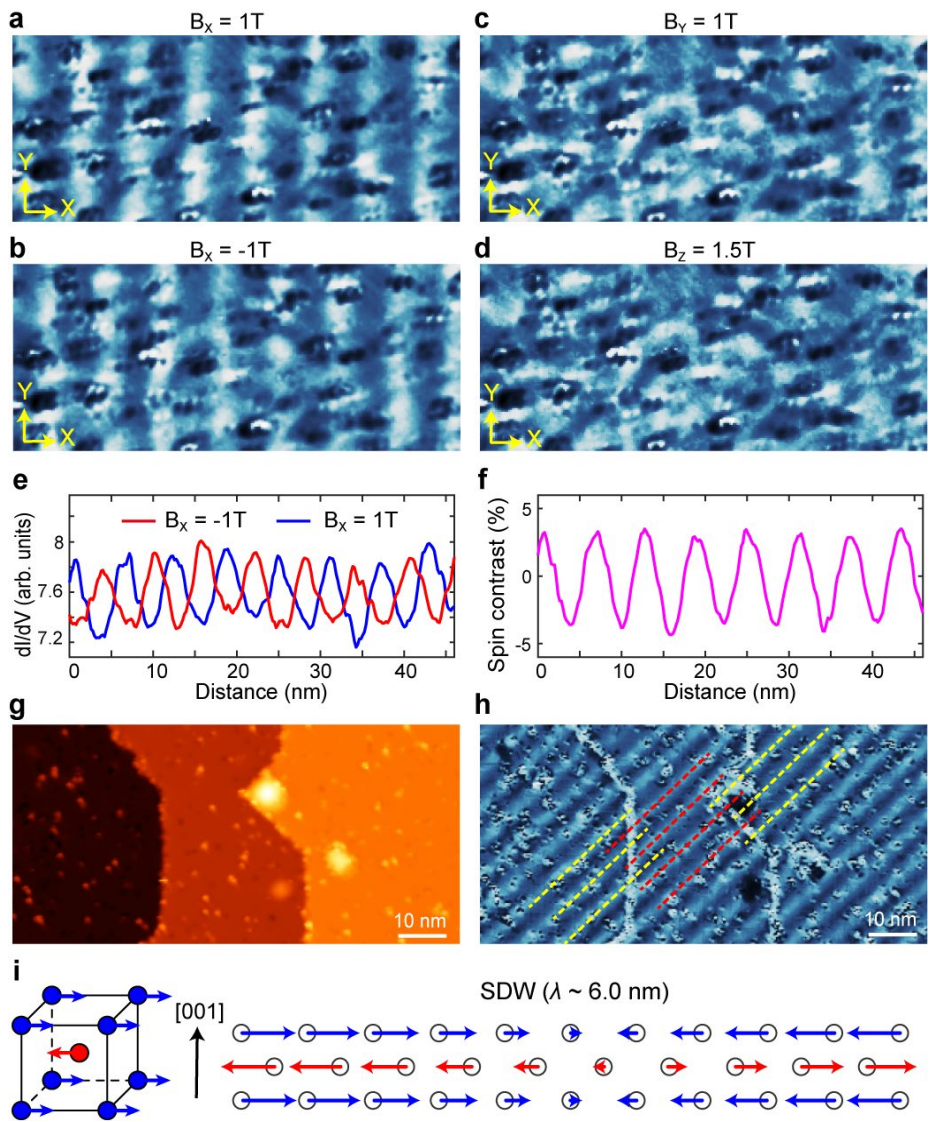

**Fig. 3 Verifying the nature of spin density wave. a-d** d$I$/d$V$ maps taken with a Fe-coated tip under different magnet field (marked above each panel). The mapping area is the same for these panels (size: 46×20 nm², $V_b$ = -100 mV, $I$ = 80 pA, $\Delta V$ = 10 mV). **e** averaged line profile (along X direction) of panels (**a**) and (**b**). A phase inversion can be clearly seen for $B_x$ = ± 1T. **f** Spin contrast (polarization ratio) calculated by the relative difference of the line profiles in panel (**e**). **g** STM image of three adjacent mono-atomic steps. The step height is ≈ 1.45 Å. **h** d$I$/d$V$ map taken in the same area of panel (**g**), the SDW displays phase inversion across each step edge, as indicated by yellow and red dashed lines ($V_b$ = -200 mV, $I$ = 100 pA, $\Delta V$ = 15 mV). **i** A sketch of local spin configuration and IC-SDW near Cr (001) surface.

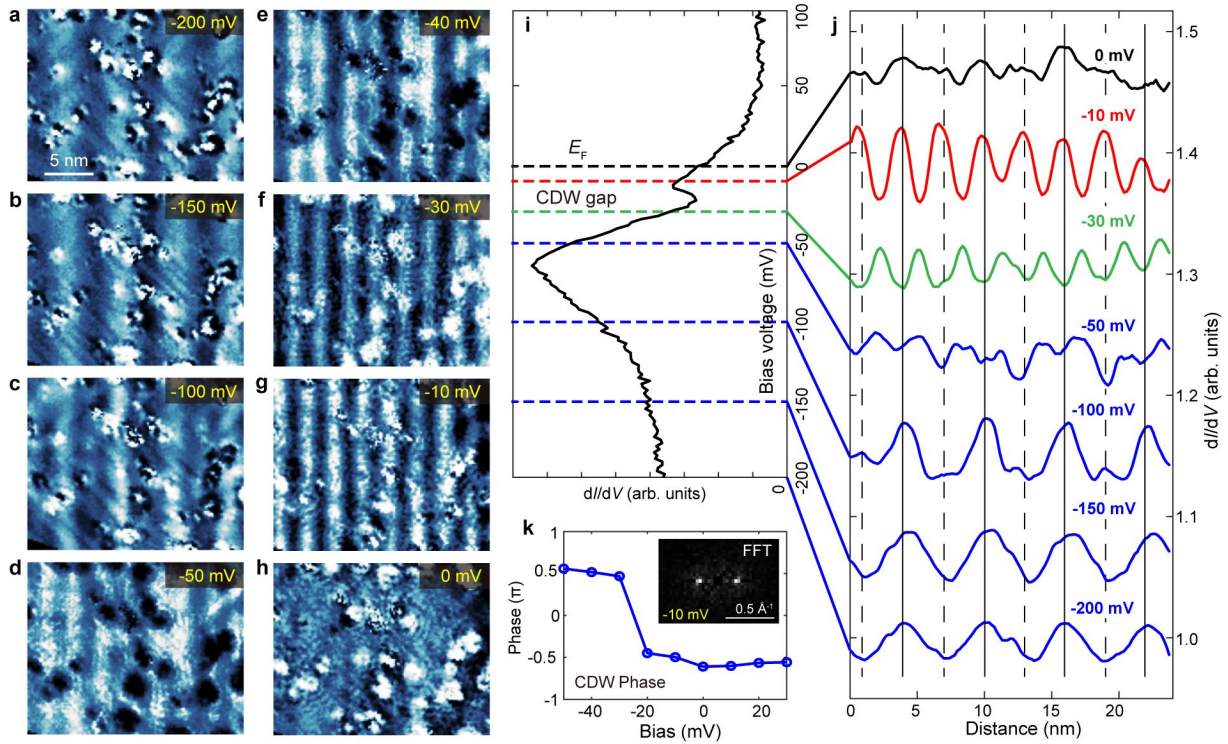

**Fig. 4 Evolvement of SDW/CDW modulations with energy, and the phase inversion of CDW. a-h** d$I$/d$V$ maps taken by Cr-coated tip at various $V_b$. The mapping area are the same for these panels. **i** The typical d$I$/d$V$ point spectrum taken in this region (setpoint: $V_b$ = -500 mV, $I$ = 60 pA, $\Delta V$ = 10 mV). **j** Averaged line profile of the d$I$/d$V$ maps. An abrupt phase inversion can be seen between $V_b$ = -30 and -10 mV, which corresponds to the DOS dip region in panel (**i**). **k** The phase of the CDW modulation, which are extracted from the raw FFT values at $\mathbf{Q}_{CDW}$. Inset image shows the FFT of panel (**g**) ($V_b$ = -10mV) which displays $\mathbf{Q}_{CDW}$ spots at ±2π/$\lambda_{CDW}$.

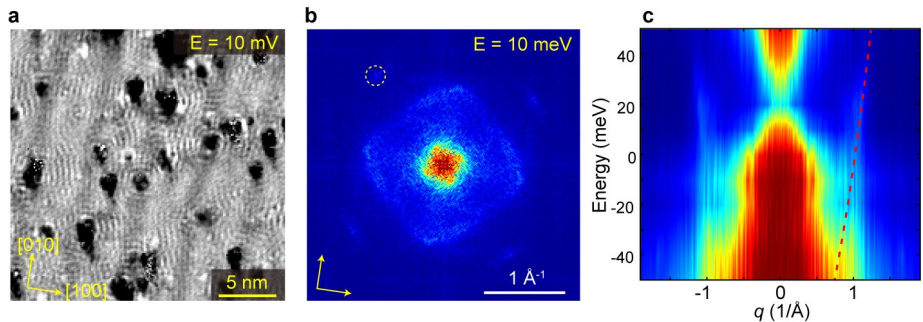

**Fig. 5 QPI measurement on Cr (001) surface. a** d$I$/d$V$ image taken at $V_b$ = 10 mV ($I$ = 100 pA), which shows the QPI modulation. **b** The four-fold symmetrized FFT image of at $E$ = 10 meV. The dashed circle indicated the FFT spots from c(2×2) reconstruction. **c** Color plot of the summarized FFT line cuts, dashed line tracks the dispersion of electron-like band.

# Supplementary materials of

## "Real-space Observation of Incommensurate Spin Density Wave and Coexisting Charge Density Wave on Cr(001) surface"

This supplementary material contains additional data of surface characterization, tunneling spectrum and d$I$/d$V$ mapping of Cr (001), included in Figs. S1-S9. Detailed descriptions are in the figure captions.

## Contains:

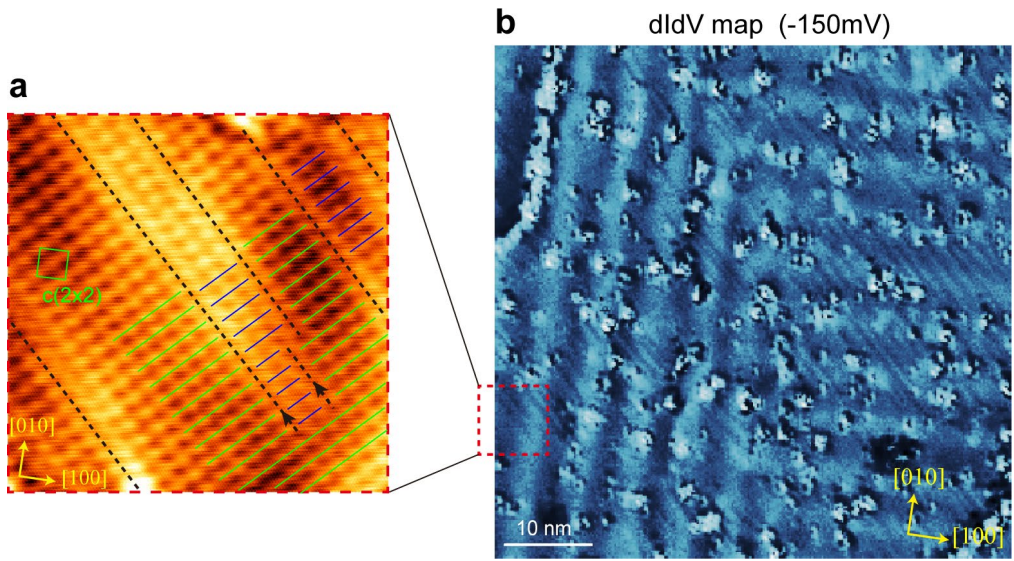

**Supplementary Figure S1. The surface dislocation lines and corresponding d$I$/d$V$ map (Cr coated tip, $T$= 5.0 K).** (**a**) Topographic image (8×8 nm$^2$) taken at the dashed box area in panel **b**. The atomic c(2×2) lattice and dislocation lines can be seen. The lattice on one side of the dislocation line gradually shifts (starting from the two arrows) with respected to the lattice on the other side, as indicated by the green/blue solid lines. (**b**) d$I$/d$V$ map taken by a Cr coated tip ($V_b$ = -150 mV, $I$ = 80 pA, $\Delta V$ = 20 mV, 60×60 nm$^2$). The short stripes along the $1\bar{1}0$ directions are caused by the dislocation lines as shown in panel **a**. They are randomly distributed with no fixed period, and they do not show influence to the long-period SDW modulation.

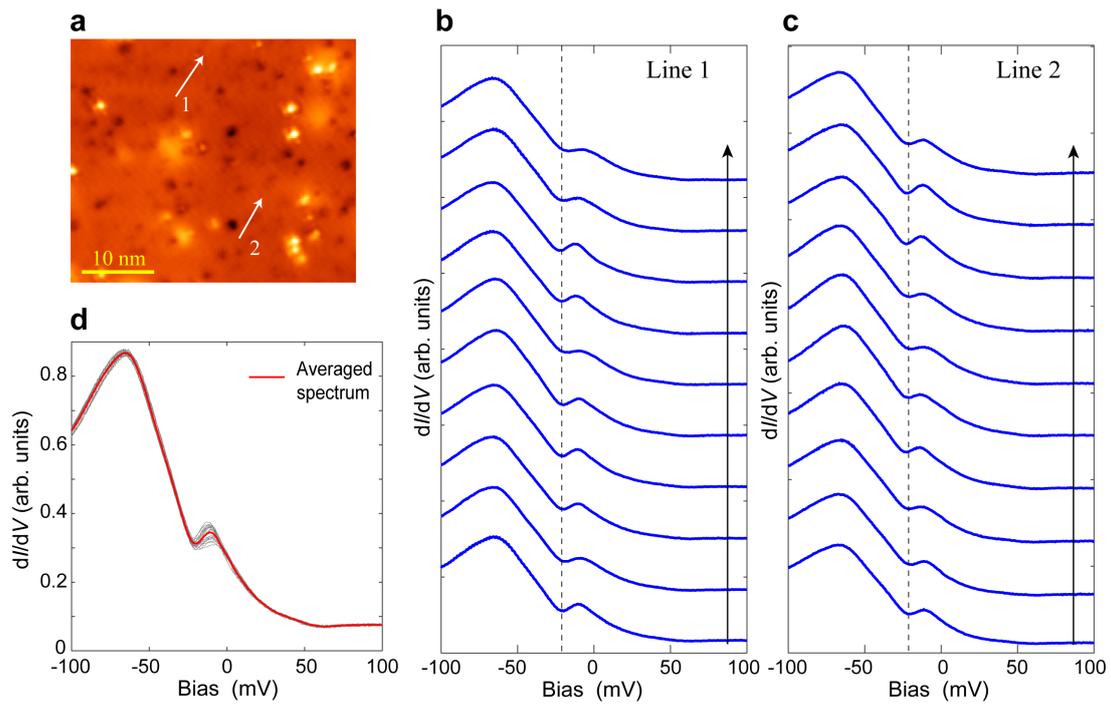

**Supplementary Figure S2. Spatial dependence of the tunneling spectrum (Pt/Ir tip, *T* = 5.0 K).** (**a**) Topographic image of Cr(001) ($V_b$ = -30 mV, *I* = 10 pA). (**b** - **c**) Series of d*I*/d*V* spectra taken along the two arrows shown in panel **a**, respectively (setpoint: $V_b$ = -100 mV, *I* = 150 pA, Δ*V* = 5 mV). The DOS dip at *E* = -22(±1) meV appears in every spectrum. The spectra are shifted vertically. (**d**) A non-shifted plot of all the spectra that shown in panel **b** and **c**, the red curve is averaged spectrum.

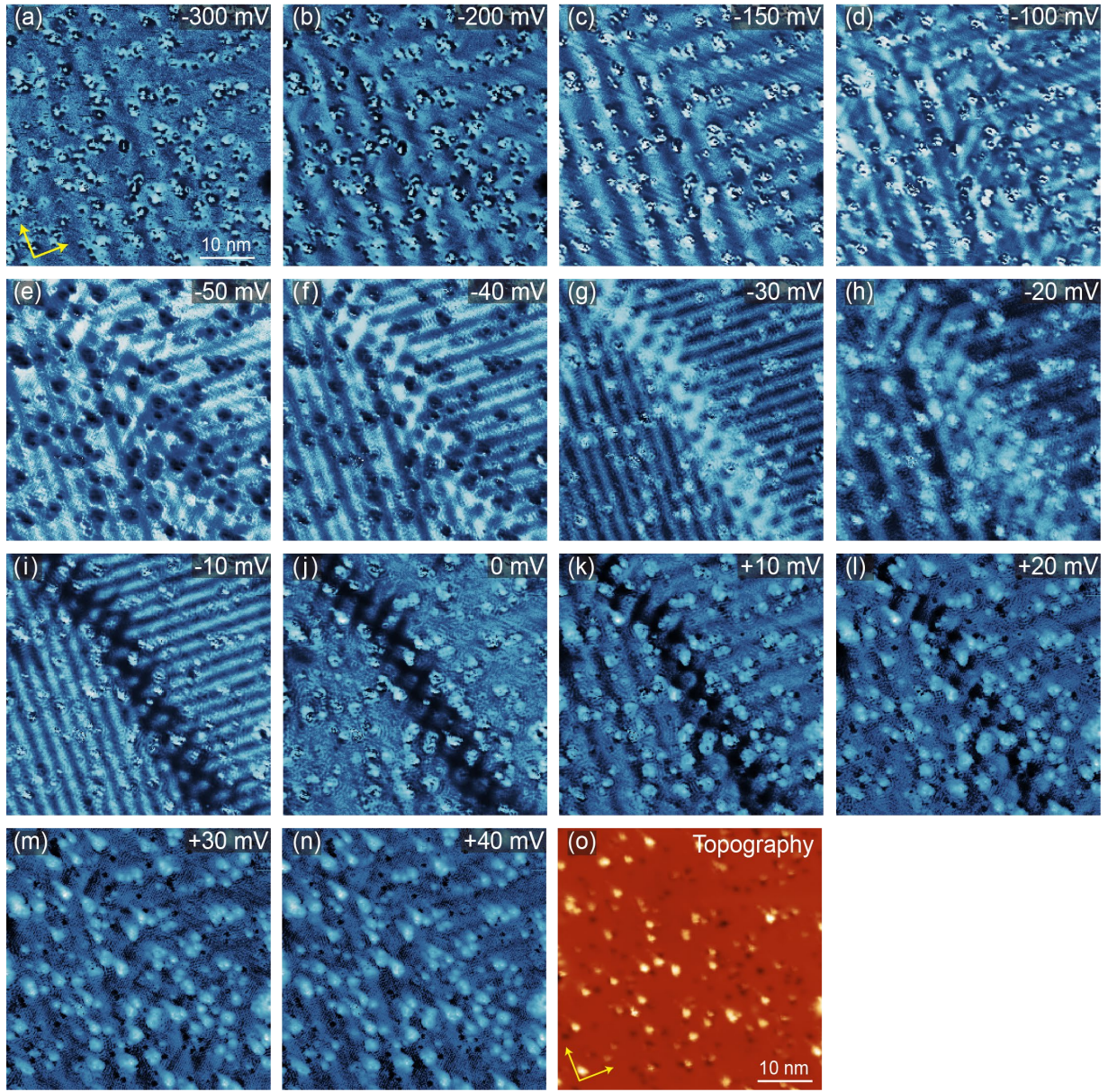

**Supplementary Figure S3.** (**a-n**) A set of d$I$/d$V$ maps taken at various energies with a Cr-coated tip ($T$ = 5.0 K). The mapping energy is labeled in each panel. (**o**) The topographic image of the mapping area. Two SDW/CDW domains can be seen in d$I$/d$V$ maps, while the domain structure is unrelated to topography. (Setpoint: $I$ = 150 pA for all the maps. $V_b$ = -300 mV for panel **a**. $V_b$ = -200 mV for panel **b**, $V_b$ = -150 mV for panel **c**, $V_b$ = -100 mV for panel **d**, $V_b$ = -50 mV for panels **e - n**. $\Delta V$ = 20 mV for panels **a - b**, $\Delta V$ = 15 mV for panel **c**, $\Delta V$ = 10 mV for panel **d** and $\Delta V$ = 5 mV for panels **e - n**.)

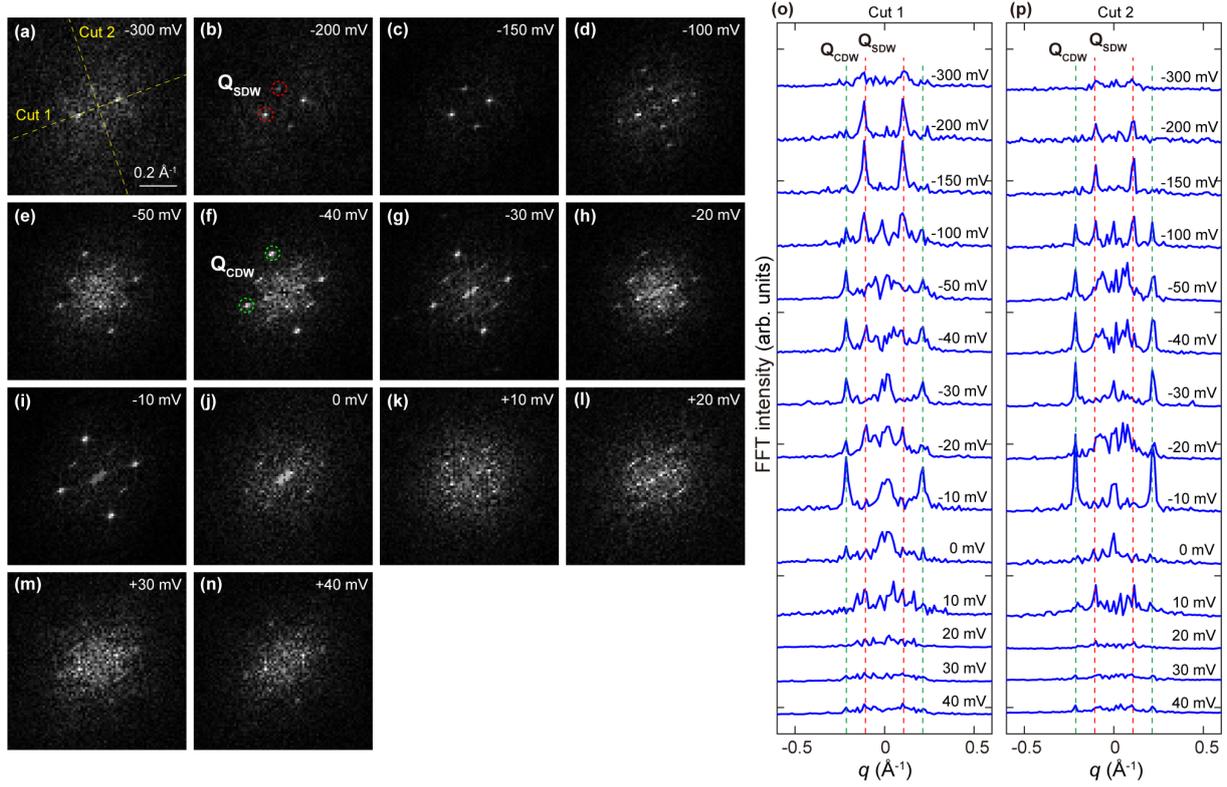

**Supplementary Figure S4. FFT images and the non-dispersive behavior of SDW/CDW.** (**a - n**) Raw FFT images of the d$I$/d$V$ maps shown in Fig. S3(a - n), respectively. Two domains of the SDW/CDW that perpendicular to each other give rise to two sets of SDW/CDW spots in FFT. The red and green dashed circles in panels **b** and **f** indicate the SDW and CDW spots, respectively. (**o - p**) FFT line-profiles at different energies, taken along the two yellow dashed lines marked in panel **a**. The green and red dashed lines indicate the positions of **Q**$_{CDW}$ = 0.21 Å$^{-1}$ and **Q**$_{SDW}$ = 0.105 Å$^{-1}$, respectively.

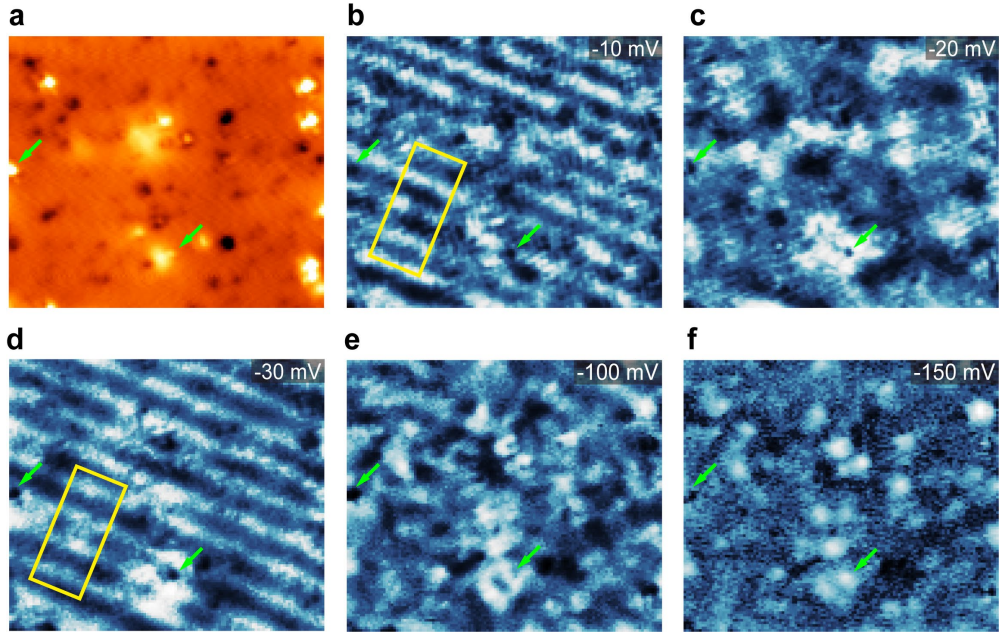

**Supplementary Figure S5. A data set measured by Pt/Ir tip at *T* = 5.0 K.** (**a**) Topographic image of Cr(001) (29×25 nm$^2$). (**b - f**) d*I*/d*V* maps taken in the same area as panel **a**. The mapping energies are labelled in each panel. The 3.0 nm period CDW modulations are still present at energies close to $E_F$, but the SDW modulations are absent in all the maps. The CDW phase inversion between -30 meV and -10 meV can also be seen (indicated by the rectangles at the same position in panels **b** and **d**). (Setpoints: **b**, *I* = 100 pA, $V_b$ = -10mV, Δ*V* = 2 mV; **c**, *I* = 100 pA, $V_b$ = -20mV, Δ*V* = 3 mV; **d**, *I* = 100 pA, $V_b$ = -30mV, Δ*V* = 2 mV; **e**, *I* = 150 pA, $V_b$ = -100mV, Δ*V* = 5 mV; **f**, *I* = 150 pA, $V_b$ = -150mV, Δ*V* = 5 mV). The two green arrows in each panel point to two defects and their induced LDOS features in d*I*/d*V* maps.

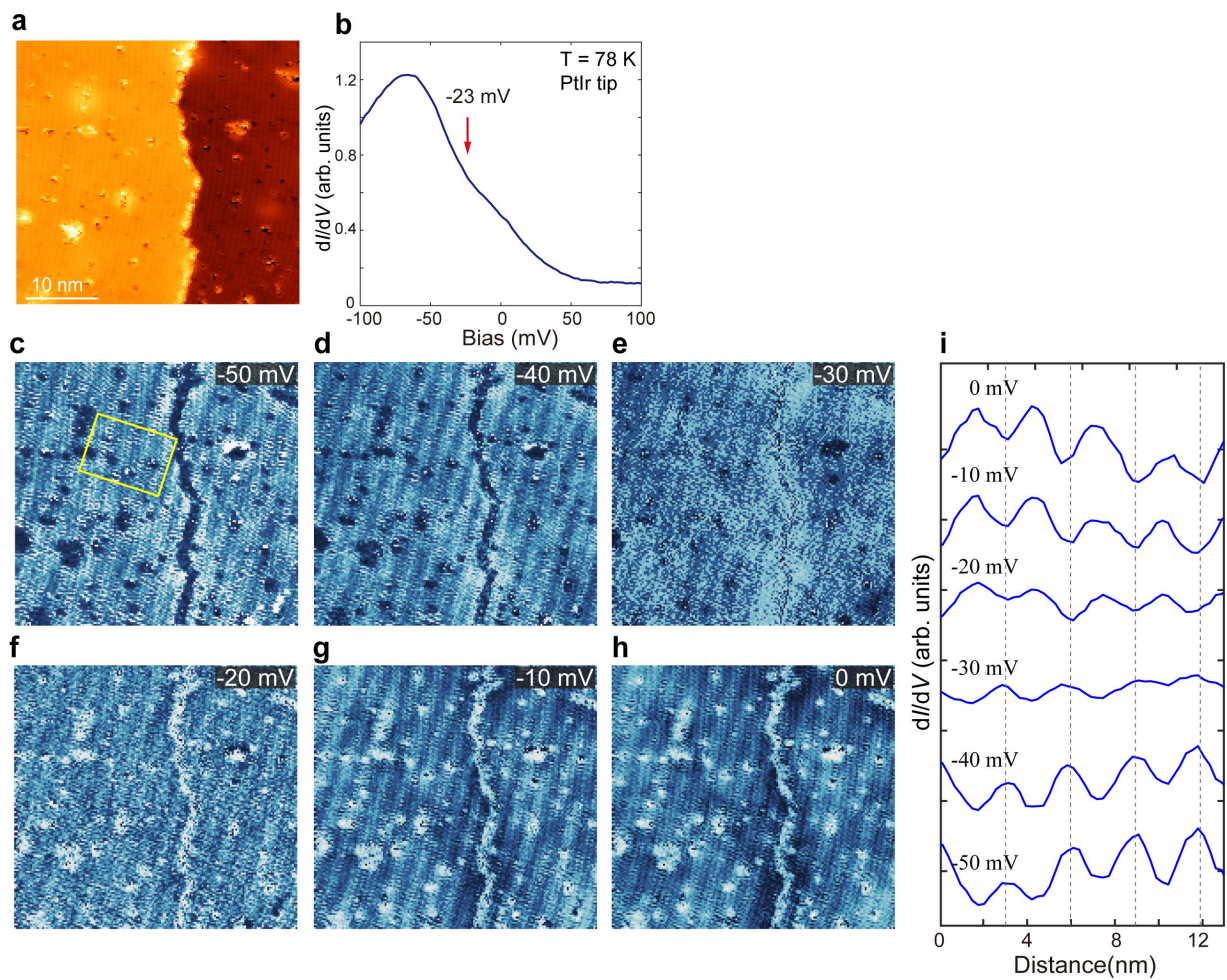

**Supplementary Figure S6. A data set measured by Pt/Ir tip at $T$ = 78 K.** (**a**) Topographic image of the Cr(001) surface ($V_b$ = -40 mV, $I$ = 50 pA). (**b**) Spatially averaged d$I$/d$V$ spectrum (setpoint: $I$ = 100 pA, $V_b$ = -100 mV, $\Delta V$ = 1 mV). A kink feature is observed around -23 meV. (**c** - **h**) d$I$/d$V$ maps taken in the same area of panel **a** and at various energies (setpoint: $I$ = 50 pA, $V_b$ = -50 mV, $\Delta V$ = 8 mV). (**i**) Line profiles of the d$I$/d$V$ maps (averaged within the region marked in panel **c**). A phase inversion can be seen between $V_b$ = -30 mV and -20 mV.

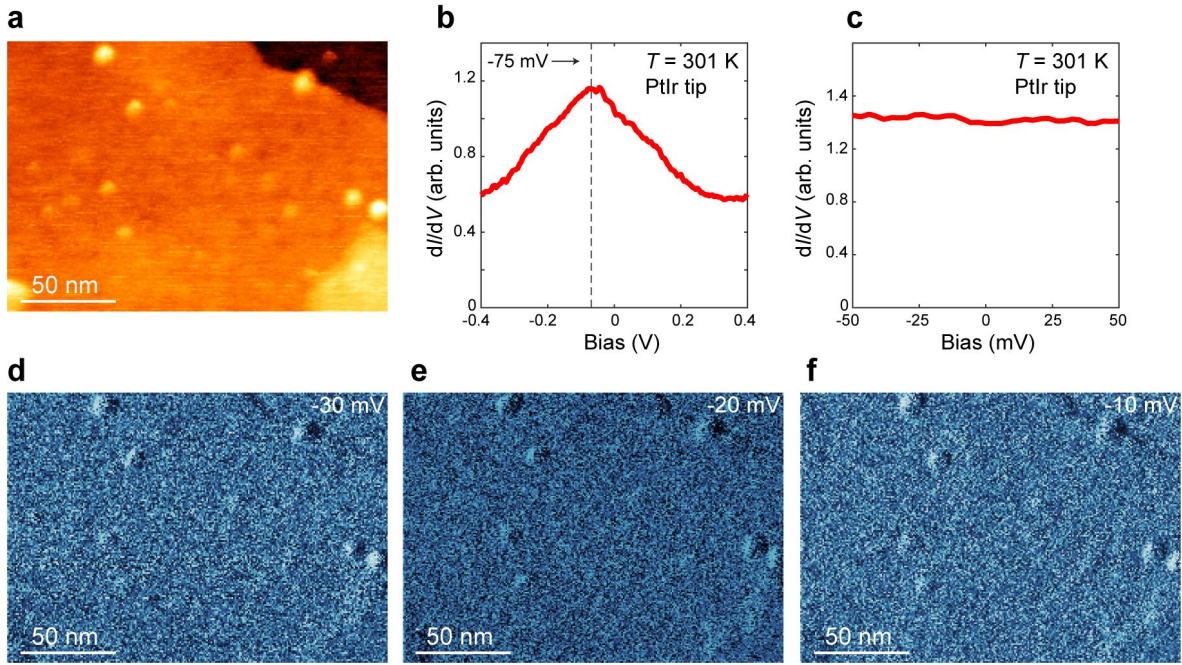

**Supplementary Figure S7. A data set measured by Pt/Ir tip at T = 301 K.** (**a**) Topographic image of Cr(001) surface ($V_b$ = -30 mV, $I$ = 100 pA). (**b - c**) Averaged d$I$/d$V$ spectra measured at defect free area. (setpoint: **b**, $I$ = 100 pA, $V_b$ = -1.0V, $\Delta V$ = 10 mV; **c**, $I$ = 100 pA, $V_b$ = -0.2V, $\Delta V$ = 1 mV). (**d - f**) Several d$I$/d$V$ maps taken in the same area as panel **a** (setpoint: $I$ = 100 pA; $V_b$ is labeled in each panel; $\Delta V$ = 3 mV). No CDW modulation is observed.

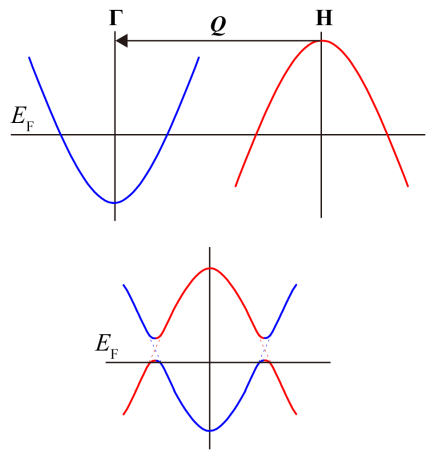

**Supplementary Figure S8.** A sketch showing the band folding from H to Γ which can induce a hybridization gap above $E_F$, as the hole pocket at H is slightly larger than the electron pocket at Γ (based on the Fermi surface of Cr reported in L. Mattheiss, Phys. Rev. 139, 1893A (1965) and D. Laurent et al., PRB 23, 4977(1981)). The folding vector between Γ and H points is **Q** = 2π/a.

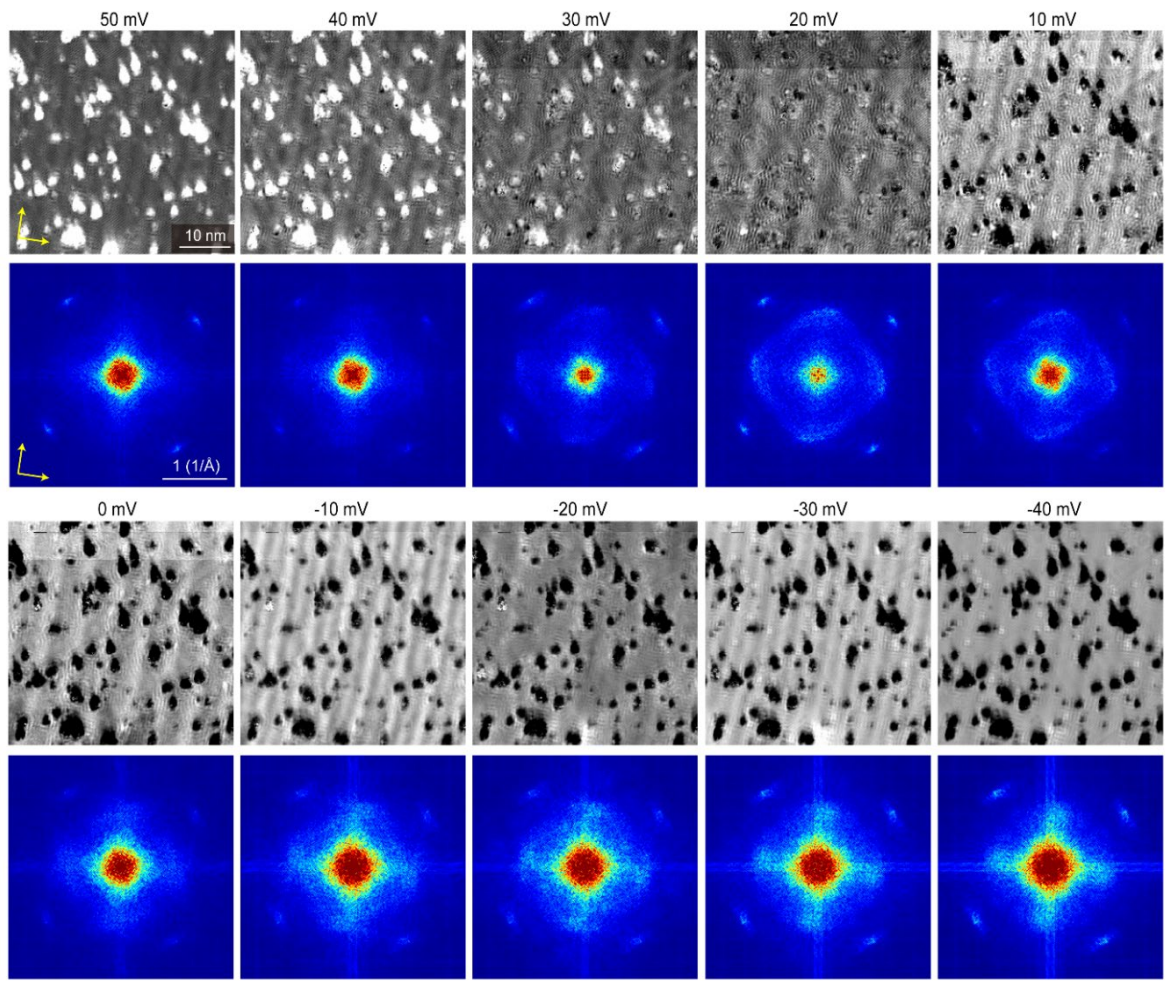

**Supplementary Figure S9. Additional data of QPI measurement:** A set of d$I$/d$V$ maps and corresponding FFT images taken by a Cr coated tip at $T$ = 5.0 K, which show QPI patterns together with the SDW/CDW modulations. The mapping energy is labeled in each panel and the FFTs are four-fold symmetrized. (Setpoint: $I$ = 100 pA, $V_b$ = 50mV, Δ$V$ = 5 mV)